\documentclass[conference]{IEEEtran}
\IEEEoverridecommandlockouts

\usepackage{algorithmic}
\usepackage{amsthm}
\usepackage{amsmath}
\usepackage{amssymb}
\usepackage{amsfonts}
\usepackage{balance}
\usepackage{caption}
\usepackage{cite}
\usepackage{color}
\usepackage{esint}
\usepackage{graphicx}
\usepackage{epstopdf}
\usepackage{subcaption}

\usepackage{epstopdf}

\usepackage{textcomp}
\def\BibTeX{{\rm B\kern-.05em{\sc i\kern-.025em b}\kern-.08em
		T\kern-.1667em\lower.7ex\hbox{E}\kern-.125emX}}

\theoremstyle{plain}
\newtheorem{thm}{\protect\theoremname}

\makeatother

\providecommand{\theoremname}{Theorem}

\theoremstyle{plain}

\makeatother

\providecommand{\lemmaname}{Lemma}

\theoremstyle{plain}

\providecommand{\propositionname}{Proposition}

\makeatother

\theoremstyle{plain}

\makeatother

\providecommand{\corname}{Corollary}

\providecommand{\remarkname}{Remark}

\theoremstyle{plain}
\newtheorem{rem}{\protect\remarkname}

\makeatother

\begin{document}
\title{Pathloss modeling of reconfigurable intelligent surface assisted THz wireless systems 
\thanks{This work has received funding from the European Commission's Horizon 2020 research and innovation programme (ARIADNE) under grant agreement No. 871464.}}

\author{\IEEEauthorblockN{ Alexandros--Apostolos A. Boulogeorgos,  and Angeliki Alexiou}
	\IEEEauthorblockA{\textit{Department of Digital Systems,
			University of Piraeus}, 
		Piraeus, Greece 
		(e-mails: al.boulogeorgos@ieee.org, alexiou@unipi.gr)}
}



\maketitle

	\begin{abstract}  
 This paper presents an analytical pathloss model for reconfigurable intelligent surface (RIS) assisted terahertz (THz) wireless systems. Specifically, the model  accommodates both the THz link and the RIS particularities. Finally, we derive a closed-form expression that returns the optimal phase shifting of each RIS reflection unit. The derived pathloss model is validated through extensive electromagnetic simulations and is expected to play a key role in the design of RIS-assisted THz wireless~systems. 
\end{abstract}



\section{Introduction}\label{S:Intro}

With our attention placed on the tremendous data traffic demands that are expected to be brought together with the sixth generation (6G) application scenarios~\cite{A:LC_CR_vs_SS,A:ED_in_FD_with_Residual_RF_impairments,ref14_Al_hard_imperf,C:Energy_Detection_under_RF_impairments_for_CR}, two technological approaches are examined as candidate solutions~\cite{Dang2020,Bariah2020,PhD:Boulogeorgos}. The first one is to move to higher-frequency bands, with emphasis on the terahertz (THz) one~\cite{Boulogeorgos2018,Rappaport2019,Boulogeorgos2019,Boulogeorgos2020a,Boulogeorgos2018a,WP:Wireless_Thz_system_architecture_for_networks_beyond_5G,C:ADistanceAndBWDependentAdaptiveModulationSchemeForTHzCommunications,C:UserAssociationInUltraDenseTHzNetworks,Boulogeorgos2019a}, while the second one is to exploit reconfigurable intelligent surfaces (RISs) capable of devising a beneficial wireless propagation environment~\cite{A:Exploration_of_intercell_wireless_millimeter_wave_communication_in_the_landscape_of_intelligent_metasurfaces,A:Smart_radio_enviroments,Boulogeorgos2021}. 

In the technical literature,  several contributions appear on analyzing, optimizing, designing, and demonstrating wireless THz systems~\cite{jornet2011,EuCAP2018_cr_ver7,Merkle2017}. All of them agree that line-of-sight (LoS) channel attenuation and blockage are the main limiting factors of THz wireless systems. To  break the barriers set by blockage, recently, some research works proposed the use of RIS~\cite{A:Performance_analysis_of_LISs,A:Reconfigurable_Intelligent_Surfaces_for_EE_in_WC,Thirumavalavan2020,Renzo2020,Bjornson2020,Boulogeorgos2020b}. In particular, in~\cite{Bariah2020}, and~\cite{A:Wireless_communications_through_RIS}, the authors explained how RIS can be used to mitigate the impact of blockage and introduced the idea of reflected LoS links. In this sense, in~\cite{A:Performance_analysis_of_LISs}, the authors conducted an asymptotic uplink ergodic capacity study, assuming that the transmitter (TX)-RIS and RIS-receiver (RX) channels follow Rician distribution. Similarly, in~\cite{A:Reconfigurable_Intelligent_Surfaces_for_EE_in_WC} the joint maximization of the sum-rate and energy efficiency was studied for a multi-user downlink scenario, in which connectivity was established by means of reflected LoS.
Additionally, in~\cite{Thirumavalavan2020}, an error analysis was performed for RIS-assisted non-orthogonal multiple access networks. Moreover, in~\cite{Renzo2020}, di Renzo et. al highlighted the fundamental similarities and differences between RISs and relays. In the same direction, in~\cite{Bjornson2020}, the authors compared the performance of RIS-assisted systems against decode-and-forward relaying ones in terms of energy efficiency, while, in~\cite{Boulogeorgos2020b}, the authors conducted a performance comparison between RIS and amplify-and-forward (AF) relays in terms of average received signal-to-noise-ratio (SNR), outage probability, diversity order and gain, symbol error rate and ergodic capacity, which revealed that, in general, RIS-assisted wireless systems can outperform the corresponding AF relaying ones.

Despite the paramount importance of combining THz wireless and RIS technologies, there are only a few published works that investigate the performance of RIS-assisted THz wireless systems~\cite{Ma2020,Qiao2020,Tekbiyik2020}. In ~\cite{Ma2020} and~\cite{Qiao2020}, although the directional nature of the THz links was taken into account, the pathloss (PL) characteristics of the transmission path were neglected, while, in~\cite{Tekbiyik2020}, the impact of molecular absorption loss was ignored.  The main reason behind this is the lack of tractable  PL model for RIS-assisted systems operating in the THz band.  
To cover this research gap, this paper focuses on providing a low-complexity PL model that takes into account the particularities of the THz propagation medium as well as the physical characteristic of the RIS. 
In more detail, the  model takes into account not only the access point (AP)-RIS and RIS-user equipment (UE) distances, but also the RIS size, the radiation pattern and the reflection coefficient of the RIS reflection unit (RU), the AP and UE antenna gain, the transmission frequency, as well as the environmental conditions\footnote{Note that there are two already published contributions that provided the end-to-end (e2e) PL in RIS-assisted wireless sytsems~\cite{Ellingson2019,Tang2019}. However, both~\cite{Ellingson2019} and~\cite{Tang2019} refer to low frequency band communications; thus, they neglect the impact of molecular absorption loss.}. 
Building upon the channel attenuation expression, we provide a closed-form expression that determines the phase shift that should be implemented on each RIS element in order to steer the reflected by the RIS beam towards the~UE.

\section{System   Model}\label{sec:SM}

\begin{figure}
	\centering
	\includegraphics[width=0.85\linewidth,trim=0 0 0 0,clip=false]{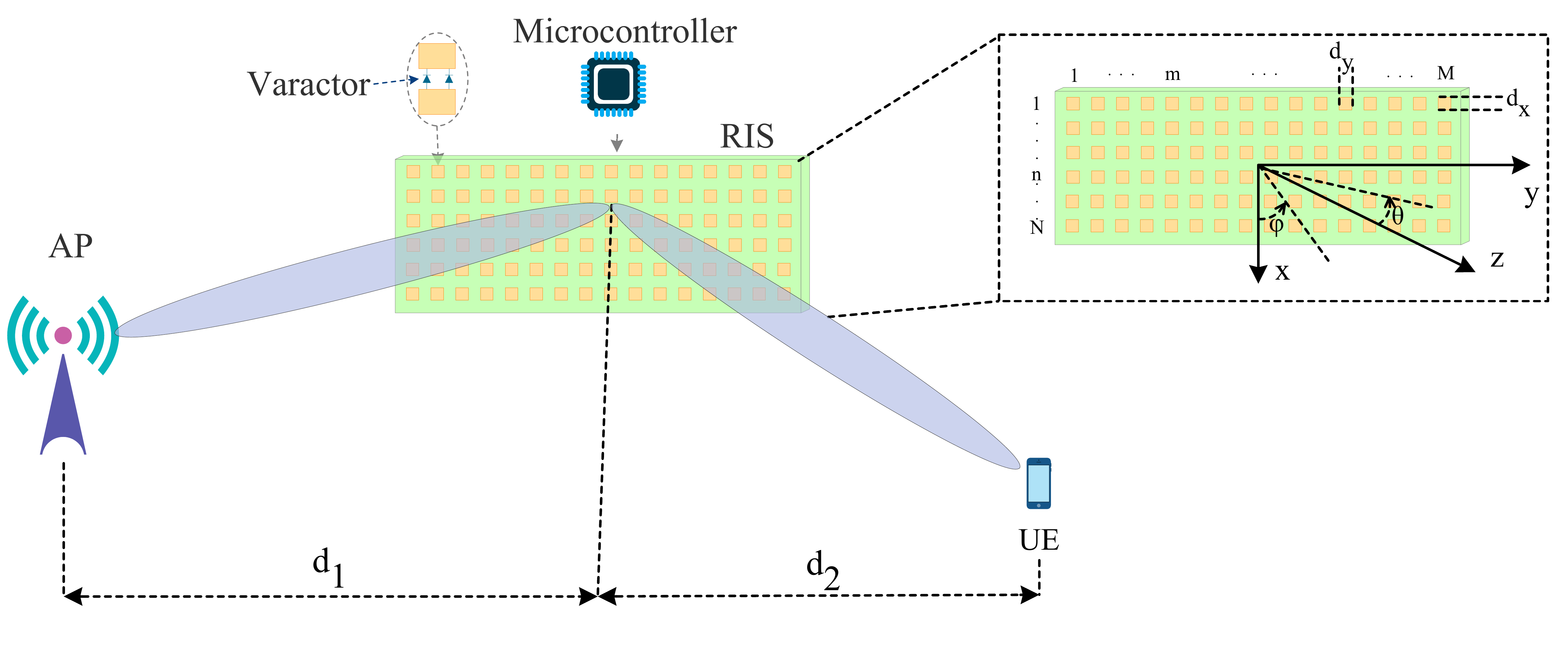}
	\caption{System model.}
	\label{fig:SM}
\end{figure}

As shown in Fig.~\ref{fig:SM}, downlink scenario of a RIS-assisted wireless THz system is considered, where a single AP serves a UE through a RIS. 
 The AP and the UE are equipped with high-directional antennas of gains $G_{a}$ and $G_{u}$, respectively. Both the AP and UE antennas are assumed to point at the center of the RIS. 
 The RIS consists of $M\times N$ orthogonal RUs of dimensions $d_x$ and $d_y$. 
 Moreover, the UE position is assumed to be known to the RIS controller.  A three dimensional (3D) Cartesian system is defined  centered at the RIS center. The RIS horizontal and vertical directions respectively define the $x$ and $y$ axis. Hence, the position of the RU, $\mathcal{U}_{m,n}$ can be obtained~as
$	\mathbf{d}_{m,n} = \left(n-\frac{1}{2}\right) d_x \mathbf{x}_o + \left(m-\frac{1}{2}\right) d_y \mathbf{y}_o + 0 \text{ } \mathbf{z}_o,
\label{Eq:d_mn}$
with $n\in\left[1-\frac{N}{2}, \frac{N}{2}\right]$ and $m\in\left[1-\frac{M}{2}, \frac{M}{2}\right]$. Also, $\mathbf{x}_o$, $\mathbf{y}_o$, and $\mathbf{z}_o$ stand for the unitary vectors at the $x$, $y$, and $z$ direction,~respectively. Finally, let $d_1$ and $d_2$ respectively denote the AP-RIS and RIS-UE distances.

\section{Path-loss model}\label{SS:CM}
Let, $\theta_{m,n}^{t}$ and $\theta_{m,n}^{r}$ be the elevation angle from the $(m, n)$ RU, $\mathcal{U}_{m,n}$, to the AP and to the UE, respectively, while $\phi_{m,n}^{t}$ and $\phi_{m,n}^{r}$ stand for the corresponding azimuth angles. Finally, we use $l_{m,n}^t$ and $l_{m,n}$ to respectively define the distances  from AP to the $\mathcal{U}_{m,n}$ RU and the one from the $\mathcal{U}_{m,n}$ RU to the UE. The following theorem returns the e2e pathloss.  
\begin{thm}\label{Thm:PL_general}
	 The e2e PL can be evaluated~as
	 in~\eqref{Eq:L_GC}, given at the top of the following page.
	\begin{figure*}
	\begin{align}
	L &= M^{-2} N^{-2} \frac{64\pi^3 d_1^2 d_{2}^2} {d_x d_y \lambda^2 |R|^2  U^{r}\left(\theta_i, \phi_i\right) U^{t}\left(\theta_r, \phi_r\right) G_{t} } 
	\frac{\mathrm{sinc}^2\left( \frac{\pi}{\lambda} \left(  \sin\left(\theta_i\right) \cos\left(\theta_i\right)+ \sin\left( \theta_r \right) \cos\left(\phi_r\right) + \zeta_1\right)  d_x \right)}{\mathrm{sinc}^2\left( \frac{N\pi}{\lambda} \left(  \sin\left(\theta_i\right) \cos\left(\theta_i\right)+ \sin\left( \theta_r \right) \cos\left(\phi_r\right) + \zeta_1 \right)  d_x \right)}
	\nonumber \\ & \times
	\frac{\mathrm{sinc}^2\left( \frac{\pi}{\lambda} \left(  \sin\left(\theta_i\right) \sin\left(\phi_i\right)+ \sin\left( \theta_r \right) \sin\left(\phi_r\right) +\zeta_2 \right)  d_y \right)}{\mathrm{sinc}^2\left( \frac{M\pi}{\lambda} \left(  \sin\left(\theta_i\right) \sin\left(\phi_i\right)+ \sin\left( \theta_r \right) \sin\left(\phi_r\right) +\zeta_2\right)  d_y \right)}
	\exp\left(\kappa(f) \left( d_1 + d_{2}\right) \right)
	\label{Eq:L_GC}
	\end{align}
	\hrulefill
	\end{figure*}
	In~\eqref{Eq:L_GC}, 
	\begin{align}
	\zeta_1 \left(n-\frac{1}{2}\right) d_x &+ \zeta_2 \left(m-\frac{1}{2}\right) d_y =  \frac{\lambda\phi_{m,n}}{2\pi},
	\label{Eq:phi_mn}
	\end{align}
	and
	\begin{align}
	G_{t} &= G_{a} G G_{u}.
	\end{align}
	Additionally, $\phi_{m,n}$ and $|R|$  are respectively the controllable phase shift and the absolute value of the reflection coefficient introduced by the $(m,n)$ RU, while  $U^{r}\left(\theta, \phi\right)$,  $U^{t}\left(\theta, \phi\right)$ and $G$  are the normalized received, the normalized transmitted power ratio patterns and the RU gain, respectively. Moreover, $\theta_i$ and $\phi_i$ are respectively the elevation and the azimuth angles from the  center of the the RIS to the AP, while $\theta_r$ and $\phi_r$ respectively denotes the the elevation and the azimuth angles from the  center of the the RIS to the center of the cluster.  
Finally, in~\eqref{Eq:L_GC}, $\kappa(f)$ stands for the molecular absorption coefficient and can be obtained as in~\cite{Kokkoniemi2020}\footnote{In practice THz wireless systems are expected to operate in the $100-500\text{ }\mathrm{GHz}$ band, we employ a simplified model for this band, which was introduced in~\cite{EuCAP2018_cr_ver7} and then extended in~\cite{Kokkoniemi2020}.}.
\begin{align}
\kappa(f) &= \sum_{i=1}^{6}
\frac{A_i\left(\mu\right)}{B_i\left(\mu\right)+\left(\frac{f}{100c}-q_i\right)^2} 
+C\left(\mu, f\right),
\label{Eq:Kappa_f}
\end{align}
where
$A_1\left(\mu\right) = a_1 \left(1-\mu\right) \left(a_2 \left(1-\mu\right) + a_3\right),$ 
$A_3\left(\mu\right) = f_1 \mu \left(f_2 \mu + f_3 \right),$
$A_2\left(\mu\right) = c_1 \mu \left(c_2 \mu + c_3 \right), $
$A_4\left(\mu\right) = i_1 \mu \left(i_2 \mu + i_3 \right),$
$A_5\left(\mu\right) = k_1 \mu \left(k_2 \mu + k_3 \right),$
$A_6\left(\mu\right) = m_1 \mu \left(m_2 \mu + m_3 \right),$
$B_1\left(\mu\right) = \left(b_1 \left(1-\mu\right) + b_2\right)^2,$
$B_2\left(\mu\right) = \left(e_1 \mu + d_2\right)^2,$
$B_3\left(\mu\right) = \left(g_1 \mu + g_2\right)^2,$ 
$B_4\left(\mu\right) = \left(j_1 \mu + j_2\right)^2, $
$B_5\left(\mu\right) = \left(l_1 \mu + l_2\right)^2, $
$B_6\left(\mu\right) = \left(n_1 \mu + n_2\right)^2,$
$C\left(\mu, f\right) = \frac{\mu}{r_1} \left(r_2 + r_3 f^{r_4}\right),$
with $a_1=5.159\times 10^{-5}$, $a_2 = - 6.65\times 10^{-5}$, $a_3=0.0159$, $b_1=-2.09\times 10^{-4}$, $b_2=0.05$, $c_1=0.1925$, $c_2=0.135$, $c_3=0.0318$, $e_1=0.4241$, $e_2=0.0998$, $f_1=0.2251$, $f_2=0.1314$, $f_3=0.0297$, $g_1=0.4127$, $g_2=0.0932$, $i_1=2.053$, $i_2=0.1717$, $i_3=0.0306$, $j_1=0.5394$, $j_2=0.0961$, $k_1=0.177$, $k_2=0.0832$, $k_3=0.0213$, $l_1=0.2615$, $l_2=0.0668$, $m_1=2.146$, $m_2=0.1206$, $m_3=0.0277$, $n_1=0.3789$, $n_2=0.0871$, $r_1=0.0157$, $r_2=2\times 10^{-4}$, $r_3=0.915\times 10^{-112}$, $r_4=9.42$, $q_1=3.96$, $q_2=6.11$, $q_3=10.84$, $q_4=12.68$, $q_5=14.65$, and $q_6=14.94$.
Moreover, $c$ is the speed of light, and $\mu$ is the volume mixing ratio of the water vapor and can be obtained~as
$\mu = p_1\left(p_2 + p_3 P\right) \exp\left(\frac{p_4\left(T-p_6\right)}{T+p_5-p_6}\right),$
where $p_1=6.1121$, $p_2=1.0007$, $p_3=3.46\times 10^{-8}$, $p_4=17.502$, $p_5=240.97\,^{o}K$, and $p_6=273.15\,^{o}K$. Furthermore, $T$ stands for the air temperature, and $P$ is the atmospheric~pressure.
\end{thm} 
\begin{IEEEproof}
	Please refer to Appendix~A. 
\end{IEEEproof}

\begin{rem}
	To steer the beam at the desired direction $\theta_r=\theta_o$ and $\phi_r=\phi_o$, the parameters $\zeta_1$ and $\zeta_2$ should be
	\begin{align}
	\zeta_1 &= -\left( \sin\left(\theta_i\right) \cos\left(\phi_i\right)+ \sin\left( \theta_o \right) \cos\left(\phi_o\right) \right)
	\end{align}
and 
	\begin{align}
	\zeta_2&= -\left(\sin\left(\theta_i\right) \sin\left(\phi_i\right)+ \sin\left( \theta_o \right) \sin\left(\phi_o\right)\right).
	\end{align} 
	In this case, based on~\eqref{Eq:phi_mn}, the phase shift of the $(m,n)$ element can be obtained as in~\eqref{Eq:phi_mn_o}, given at the top of the following page.
	\begin{figure*}
	\begin{align}
	\phi_{m,n}^{o}=& -\frac{2\pi}{\lambda}\left(n-\frac{1}{2}\right) \left( \sin\left(\theta_i\right) \cos\left(\theta_i\right)+ \sin\left( \theta_o \right) \cos\left(\phi_o\right) \right) d_x 
	- \frac{2\pi}{\lambda} \left(m-\frac{1}{2}\right) \left(\sin\left(\theta_i\right) \sin\left(\theta_i\right)+ \sin\left( \theta_o \right) \sin\left(\phi_o\right)\right) d_y
	\label{Eq:phi_mn_o}
	\end{align}
	\hrulefill
	\end{figure*}
	In this case, according to~\eqref{Eq:L_GC}, the minimum PL~is
	\begin{align}
		L_{o} &=  M^{-2} N^{-2} \frac{64\pi^3 d_1^2 d_{n_u}^2}{d_x d_y \lambda^2 |R|^2  U^{r}\left(\theta_o, \phi_o\right) U^{t}\left(\theta_o, \phi_o\right) G_{t} }  
		\nonumber \\ & \times
		\exp\left(\kappa(f) \left( d_1 + d_{n_u}\right) \right).
		\label{Eq:L_n_u_max}
	\end{align}
\end{rem}


\section{Numerical Results \& Discussion}\label{sec:Results}

In this section, we present numerical results, which verify the accuracy of the PL model and highlight the propagation characteristics of RIS-assisted THz wireless systems. In this direction, unless otherwise stated, we investigate the following insightful scenario. We consider standard environmental conditions, i.e.,  relative humidity $50\%$, atmospheric pressure  $101325\text{ }\mathrm{Pa}$, and temperature $296^{o}\mathrm{K}$. The AP transmission antenna gain is $50\text{ }\mathrm{dBi}$, which, according to~\cite{Boulogeorgos2020,A:Wireless_Sub_THz_Communication_System_With_High_Data_Rate,A:Advances_in_THz_communications_accelerated_by_photonics}, is a realistic value for THz wireless systems, while the UE received antenna gains are $20\text{ }\mathrm{dBi}$. The antenna pattern of the RUs is described by~\cite{Stutzman2013} 
\begin{align}
U\left(\theta, \phi\right) = \left\{
\begin{array}{l l} 
\cos\left(\theta\right), & \theta\in[0, \frac{\pi}{2}] \text{ and } \phi\in[0, 2\pi]\\
0, & \text{otherwise}.
\end{array}
\right.
\label{Eq:U}
\end{align}   
Thus, $G$ can be obtained~as
$G =\int_{0}^{2\pi} \int_{0}^{\frac{\pi}{2}} U\left(\theta, \phi\right) \sin\left(\theta \right) \text{ }\mathrm{d\theta} \text{ } \mathrm{d\phi},$
which by substituting~\eqref{Eq:U} and performing the integration returns~$G=4$. Moreover, $|R|$ is set to $0.9$, which is in-line with~\cite{Asadchy2016}. Finally, note that, in what follows, we use continuous lines and markers to respectively denote theoretical and simulation results.

\begin{figure}
	\centering
	\includegraphics[width=0.7\linewidth,trim=0 0 0 0,clip=false]{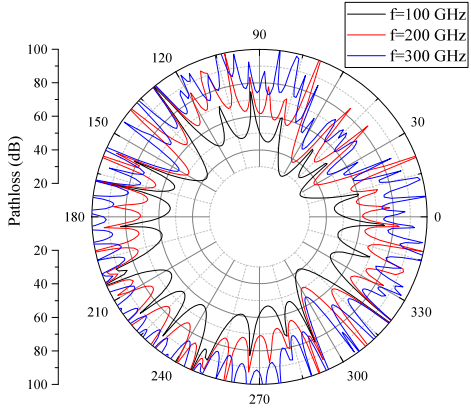}
	\caption{PL vs $\phi_r$, for $\theta_r=\frac{\pi}{4}$, $\theta_o=\pi/6$, $\phi_o=\pi/3$, and different values of $f$.}
	\label{fig:PL_vs_phir_f}
\end{figure}  

In Fig.~\ref{fig:PL_vs_phir_f}, the PL is depicted as a function of $\phi_t$, for different transmission frequencies, assuming that  $d_1=d_{2}=1\text{ }\mathrm{m}$, $d_x=d_y=0.3\text{ }\mathrm{mm}$, $|R|=0.9$, $G_{AP}=50\text{ }\mathrm{dBi}$, $G_{nu}=20\text{ }\mathrm{dBi}$, $\theta_i=\frac{\pi}{4}$, $\phi_i=\pi$, $\theta_o=\pi/6$, and $\phi_o=\pi/3$. As expected the minimum PL is observed for $\phi_r=\pi/3$. Moreover, it is apparent that for a fixed $\phi_r$, as the transmission frequency increases, the PL also increases. Finally, we observe that as the transmission frequency increases, the azimuth half power beamwidth decreases.  

\begin{figure}
	\centering
		\includegraphics[width=0.7\linewidth,trim=0 0 0 0,clip=false]{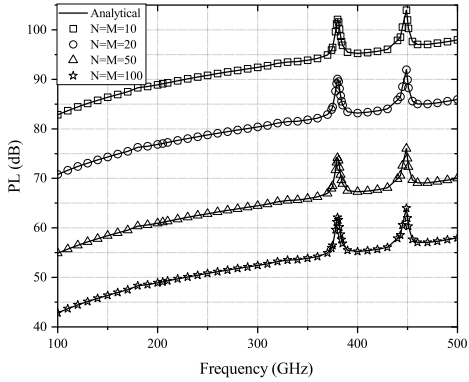}
		\caption{PL vs $f$ for different RIS sizes.}
		\label{fig:PL_vs_f}
\end{figure}   

Figure~\ref{fig:PL_vs_f} illustrates the PL as a function of $f$ for different values of $M=N$, assuming that  $\theta_i=\theta_r=\theta_o=\phi_r=\phi_o=\frac{\pi}{4}$, $\phi_i=\frac{3\pi}{4}$, $d_1=d_{2}=10\text{ }\mathrm{m}$,  and $d_x=d_y=0.3\text{ }\mathrm{mm}$. From this figure, it is revealed that there exists two frequency regions, the first one from $370$ to $390\text{ }\mathrm{GHz}$ and the second one from $430$ to $455\text{ }\mathrm{GHz}$, in which the PL is maximized. This is due to water molecules resonance. In other words, from $100$ to $500\text{ }\mathrm{GHz}$, there exists three transmission windows; the first one from $100$ to $365\text{ }\mathrm{GHz}$, the second one from $375$ to approximately $430\text{ }\mathrm{GHz}$, and the third one from $460$ to $500\text{ }\mathrm{GHz}$.
Outside these regions, for fixed $M$ and $N$, as the transmission frequency increases, the PL also increases. For example, for $M=N=20$, as $f$ increases from $100$ to $300\text{ }\mathrm{GHz}$, the PL increases for about $10\text{ }\mathrm{dB}$. Finally, it is observed that, for a given transmission frequency, as the RIS size increases, the PL decreases.For example, as $M=N$ increases from $10$ to $100$, the PL decreases for about $40\text{ }\mathrm{dB}$.

\begin{figure}
	\centering
	\includegraphics[width=0.75\linewidth,trim=0 0 0 0,clip=false]{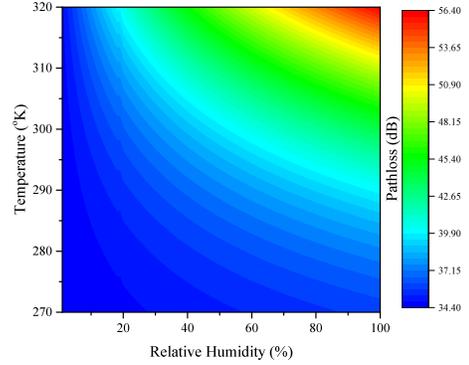}
	\caption{PL vs temperature and relative humidity.}
	\label{fig:PL_vs_T_phi}
\end{figure}  

In Fig.~\ref{fig:PL_vs_T_phi}, the PL is plotted as a function of the atmospheric temperature and relative humidity, assuming $M=N=100$, $f=380\text{ }\mathrm{GHz}$, $d_1=1\text{ }\mathrm{m}$, $d_{2}=10\text{ }\mathrm{m}$, $d_x=d_y\approx0.3\text{ }\mathrm{mm}$, $\theta_i=45^o$, $\phi_i=180^o$, and $\theta_r=\theta_o=\phi_r=\phi_o=45^o$. 
As expected, for a fixed atmospheric temperature, as the relative humidity increases, the water molecules' density increases; as a consequence, the molecular absorption and the PL increase. For instance, for $T=273^{o}K$, the PL increases by approximately $2\text{ }\mathrm{dB}$ as the relative humidity increases from $10\%$ to $90\%$. Similarly, for a given relative humidity, as the atmospheric temperature increases, the PL also increases. For example, for a $50\%$ relative humidity, the PL increases by $2\text{ }\mathrm{dB}$ as the atmospheric temperature increase from $270^oK$ to $290^oK$. Finally, by taking into account that neglecting the molecular absorption loss would lead to a PL approximately equal to $34.4\text{ }\mathrm{dB}$, it becomes evident that in this case the PL computation error could exceed $20\text{ }\mathrm{dB}$. This indicates the importance of taking into account the molecular absorption loss when evaluating the PL and the performance of RIS-assisted THz systems.

\begin{figure}
	\centering
	\includegraphics[width=0.75\linewidth,trim=0 0 0 0,clip=false]{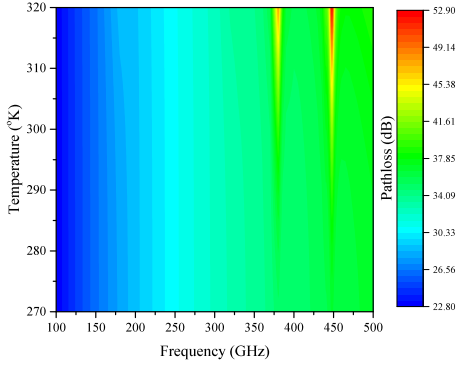}
	\caption{PL vs temperature and frequency.}
	\label{fig:PL_vs_T_fr}
\end{figure} 

Figure~\ref{fig:PL_vs_T_fr} illustrates the PL as a function of the air temperature and the transmission frequency, assuming $M=N=100$, $d_1=1\text{ }\mathrm{m}$, $d_{2}=10\text{ }\mathrm{m}$, $d_x=d_y\approx0.3\text{ }\mathrm{mm}$, $\theta_i=45^o$, $\phi_i=180^o$, and $\theta_r=\theta_o=\phi_r=\phi_o=45^o$. As expected, for a given transmission frequency, as the air temperature increases, the PL also increases. For example, for $f=250\text{ }\mathrm{GHz}$, the PL increases by about $0.1\text{ }\mathrm{dB}$, as the air temperature increases from $270$ to $320^oK$.  Moreover, from this figure, it is verified that there exist two frequency regions in which the PL is maximized. In these regions, temperature variations cause a more severe impact on PL. For instance, increasing the air temperature from $270$ to $280^oK$ results in $0.02\text{ }\mathrm{dB}$ PL increase, if $f=280\text{ }\mathrm{GHz}$, while, the same temperature increase cause a $0.5\text{ }\mathrm{dB}$ PL increase, when $f=383\text{ }\mathrm{GHz}$. This indicates the importance of taking into account the air-temperature and its variations, when selecting the transmission~frequency.

\section{Conclusions} \label{sec:Conclusions}

In this paper, we described the system model and we employed electromagnetic theory tools in order to extract a generalized formula for the e2e PL. This formula revealed the relationships between the RIS specifications, namely size, number of RIS RUs, RU size and reflection coefficient, RU's radiation patterns, as well as phase shift of each RU, the transmission parameters, such as transmission frequency AP to center of RIS and center of RIS to UE distance, AP transmission and UE reception antenna gains, azimuth and elevation angles from the AP to the center of the RIS as well as from the center of the RIS to the UE, and THz-specific parameters, like the environmental conditions that affect the molecular absorption. Building upon this expression, we determined the optimal phase shift of each RU in order to steer the RIS-generated beam to a desired direction. This work is expected to contribute on analyzing, simulating, and designing RIS-assisted THz systems.

\section*{Appendix}

\section*{Proof of Theorem 1} 

As $l_{m,n}^{t}>> \lambda$, where $\lambda$ is the wavelength of the transmission signal, the  the incident signal power at $\mathcal{U}_{m,n}$ can be obtained~as
\begin{align}
P_{m,n}^{i} &= 
\frac{G_{AP} U^{r}\left(\theta_{m,n}^{t}, \phi_{m,n}^{t}\right) d_x d_y P_{AP}}{4\pi \left(l_{m,n}^{t}\right)^2} \exp\left(-\kappa(f) l_{m,n}^{t}\right) .
\label{Eq:Pmn_i_s2}
\end{align}
Thus, the incident signal's electric field at $\mathcal{U}_{m,n}$ can be written~as
\begin{align}
	E_{n,m}^{i} = \sqrt{\frac{2 Z_o P_{m,n}^{i}}{d_x d_y}} \exp\left(-j \frac{2\pi l_{m,n}^{t}}{\lambda}\right),
\end{align} 
where $Z_o$ is the air characteristic impedance.

The total reflected signal power by $\mathcal{U}_{mn}$ can be obtained~as
$P_{m,n}^{r} = |R_{m,n}|^2  P_{m,n}^{i},$
or 
\begin{align}
P_{m,n}^{r} &=   \exp\left(-\kappa(f) l_{m,n}^{t}\right) \frac{|R_{m,n}|^2 d_x d_y}{4\pi \left(l_{m,n}^{t}\right)^2}  
\nonumber \\ & \times
 U^{r}\left(\theta_{m,n,n_u}^{t}, \phi_{m,n,n_u}^{t}\right) G_{AP} P_{AP}.
 \label{Eq:Pmn_r}
\end{align}	
By assuming that $l_{m,n}>>\lambda$, we can obtain the received signal power at the UE from $\mathcal{U}_{m,n}$~as
\begin{align}
P_{m,n} &= \exp\left(-\kappa(f)\left( l_{m,n}^{t} + l_{m,n}\right)\right)
\nonumber \\ & \times
 \frac{G\text{ } U^{r}\left(\theta_i, \phi_i\right) P_{m,n}^{r}}{4 \pi \left(l_{m,n}\right)^2} 
U^{t}\left(\theta_r, \phi_r\right) S_{r},
\label{Eq:P_m_n_nu}
\end{align}
where 
	$ S_{r} = \frac{G_{u} \lambda^2}{4\pi}$
is the aperture of the  UE receive antenna. 
Thus, the electrical field of the received signal at the UE from  $\mathcal{U}_{m,n}$ can be expressed~as 
\begin{align}
	E_{m,n} =\sqrt{2 Z_o \frac{P_{m,n}}{S_{r}}} \exp\left(-j\frac{2\pi}{\lambda}\left( l_{m,n}^{t} + l_{m,n}\right) \right),
\end{align}
or
\begin{align}
	E_{m,n} &= \frac{R_{m,n}\sqrt{{ 2 Z_o d_x d_y  U^{r}\left(\theta_i, \phi_i\right) U^{t}\left(\theta_r, \phi_r\right) G  G_{AP} P_{AP}}}}{4 \pi l_{m,n} l_{m,n}^{t}} 
	\nonumber \\ & \times
	\exp\left(-\left(\frac{1}{2}\kappa(f) + j\frac{2\pi}{\lambda} \right)\left( l_{m,n}^{t} + l_{m,n}\right) \right).
	\label{E_mnnu_s2}
\end{align} 
Hence, by taking into account that $\left|R_{m,n}\right|\approx \left|R\right|$, the total electric field at the UE can be evaluated~as
 in~\eqref{Eq:Enu}, given at the top of this page.
\begin{figure*}
\begin{align}
E_{r} & = \frac{|R|\sqrt{{ 2 Z_o d_x d_y U^{r}\left(\theta_i, \phi_i\right) U^{t}\left(\theta_r, \phi_r\right) G  G_{AP} P_{AP}}}}{4 \pi } 
\hspace{-0.3cm}
\sum_{m=-\frac{M}{2}+1}^{\frac{M}{2}} \sum_{n=-\frac{N}{2}+1}^{\frac{N}{2}} \frac{\exp\left(-\left(\frac{1}{2}\kappa(f) + j\frac{2\pi}{\lambda} \right)\left( l_{m,n}^{t} + l_{m,n}\right) + j \phi_{m,n} \right)}{l_{m,n} l_{m,n}^{t}}
\label{Eq:Enu}
\end{align}
\hrulefill
\end{figure*}

The AP position can be obtained~as
\begin{align}
\mathbf{r}_t &= d_1 \sin\left(\theta_i\right) \cos\left(\phi_i\right) \mathbf{x}_o + d_1  \sin\left(\theta_i\right) \sin\left(\phi_i\right) \mathbf{y}_o 
\nonumber  \\ & 
+ d_1 \cos\left(\theta_i\right) \mathbf{z}_o.
\label{Eq:r_t} 
\end{align}
By combining~\eqref{Eq:r_t} with the AP-$\mathcal{U}_{m,n}$ distance expression, applying the Taylor expansion in the resulting expression and keeping only the first term, the distance between the AP and the $\mathcal{U}_{m,n}$ can be approximated~as
\begin{align}
	l_{m,n}^{t}& \approx d_1 - \sin\left( \theta_i \right) cos\left(\phi_i\right) \left(n-\frac{1}{2}\right) d_x 
	\nonumber \\ & 
	- \sin\left( \theta_i \right) sin\left(\phi_i\right) \left(m-\frac{1}{2}\right) d_y
	\label{Eq:lmnt}  
\end{align} 
Following the same steps, we prove that
\begin{align}
	l_{m,n}\approx & d_{2} - \sin\left( \theta_r \right) cos\left(\phi_r\right) \left(n-\frac{1}{2}\right) d_x 
	\nonumber \\ & 
	- \sin\left( \theta_r \right) sin\left(\phi_r\right) \left(m-\frac{1}{2}\right) d_y.
	\label{Eq:lmnnu}  
\end{align}

By substituting~\eqref{Eq:lmnt} and~\eqref{Eq:lmnnu} into~\eqref{Eq:Enu}, and taking into account that in practice $d_x$ and $d_y$ are at the order of $\lambda/10$, while $d_1, d_{2}>>\lambda$, we can tightly approximate the electric field at the UE~as in~\eqref{Eq:E_r_n_u}, given at the top of the following page.
\begin{figure*} 
\begin{align}
	E_{r} & \approx \frac{|R|\sqrt{{ 2 Z_o d_x d_y  U^{r}\left(\theta_i, \phi_i\right) U^{t}\left(\theta_r, \phi_r\right) G G_{AP} P_{AP}}}}{4 \pi d_1 d_{2} } \exp\left(-\frac{1}{2}\kappa(f) \left( d_1 + d_{n_u}\right) \right)
	\nonumber \\ & \times 
	\sum_{m=-\frac{M}{2}+1}^{\frac{M}{2}} \sum_{n=-\frac{N}{2}+1}^{\frac{N}{2}}  \exp\left( j\frac{2\pi}{\lambda} \left( d_1 + d_{2}-\beta+\frac{\lambda}{2\pi}\phi_{m,n}\right) \right)
	\label{Eq:E_r_n_u}
\end{align} 
\hrulefill
\end{figure*}
Of note, in~\eqref{Eq:E_r_n_u}, 
\begin{align}
	\beta &= d_1 - \sin\left(\theta_i\right) \cos\left(\theta_i\right) \left(n-\frac{1}{2}\right) d_x 
	\nonumber \\ & 
	-  \sin\left(\theta_i\right) \sin\left(\theta_i\right) \left(m-\frac{1}{2}\right) d_y 
	\nonumber \\ & 
	+ d_{2} - \sin\left( \theta_r \right) cos\left(\phi_r\right) \left(n-\frac{1}{2}\right) d_x 
	\nonumber \\ & 
	- \sin\left( \theta_r \right) sin\left(\phi_r\right) \left(m-\frac{1}{2}\right) d_y.
\end{align}
The received signal power at UE can be evaluated~as 
\begin{align}
	P_r = \frac{|E_{n_u}^r|^2}{2 Z_o} S_{r},
\end{align}
which, with the aid of~\eqref{Eq:E_r_n_u}, can be written~as 
\begin{align}
	P_r &= \frac{d_x d_y \lambda^2 |R|^2  U^{r}\left(\theta_t, \phi_t\right) U^{t}\left(\theta_r, \phi_r\right) G G_{AP} G_{n_u} P_{AP} }{64\pi^3 d_1^2 d_{2}^2} 
	\nonumber \\ & \hspace{+2.3cm} \times 
	\exp\left(-\kappa(f) \left( d_1 + d_{2}\right) \right) \left|\gamma\right|^2.
	\label{Eq:Pr}
\end{align}
In~\eqref{Eq:Pr}, 
\begin{align}
\gamma &=  \gamma_1 \gamma_2,
\label{Eq:gamma_s2}
\end{align}
where $\gamma_1$ and $\gamma_2$ can respectively be obtained as in~\eqref{Eq:gamma1} and~\eqref{Eq:gamma_2_s2}, given at the top of the following page. 
\begin{figure*}
\begin{align}
\gamma_1 = \sum_{n=-\frac{N}{2}+1}^{\frac{N}{2}} 
\exp\left(j\frac{2\pi}{\lambda} \left(  \sin\left(\theta_i\right) \cos\left(\theta_i\right) \left(n-\frac{1}{2}\right) d_x + \sin\left( \theta_r \right) cos\left(\phi_r\right) \left(n-\frac{1}{2}\right) d_x  + \zeta_1 \right) \right)
\label{Eq:gamma1}
\end{align}
\hrulefill
\end{figure*} 
\begin{figure*}
\begin{align}
\gamma_2 = \sum_{m=-\frac{M}{2}+1}^{\frac{M}{2}}\exp\left(j\frac{2\pi}{\lambda} \left(  \sin\left(\theta_i\right) \sin\left(\theta_i\right) \left(m-\frac{1}{2}\right) d_y+ \sin\left( \theta_r \right) sin\left(\phi_r\right) \left(m-\frac{1}{2}\right) d_y +\zeta_2\right) \right)
\label{Eq:gamma_2_s2}
\end{align}
\hrulefill
\end{figure*} 
In~\eqref{Eq:gamma1} and~\eqref{Eq:gamma_2_s2}, $\zeta_1$ and $\zeta_2$ are defined in~\eqref{Eq:phi_mn}.
By taking into account the sum of geometric progression~theorem, and after performing some simple mathematical manipulations, \eqref{Eq:gamma1} can be rewritten~as
\begin{align}
\gamma_1 = N \frac{\mathrm{sinc}\left( \frac{N\pi}{\lambda} \left(  \sin\left(\theta_i\right) \cos\left(\phi_i\right)+ \sin\left( \theta_r \right) \cos\left(\phi_r\right) +\zeta_1\right)  d_x \right)}{\mathrm{sinc}\left( \frac{\pi}{\lambda} \left(  \sin\left(\theta_i\right) \cos\left(\phi_i\right)+ \sin\left( \theta_r \right) \cos\left(\phi_r\right) +\zeta_1\right)  d_x \right)}.	
\label{Eq:gamma_1_s5}
\end{align}
Similarly,~\eqref{Eq:gamma_2_s2} can be expressed~as
\begin{align}
\gamma_2 = M \frac{\mathrm{sinc}\left( \frac{M\pi}{\lambda} \left(  \sin\left(\theta_i\right) \sin\left(\phi_i\right)+ \sin\left( \theta_r \right) \sin\left(\phi_r\right) +\zeta_2\right)  d_y \right)}{\mathrm{sinc}\left( \frac{\pi}{\lambda} \left(  \sin\left(\theta_i\right) \sin\left(\phi_i\right)+ \sin\left( \theta_r \right) \sin\left(\phi_r\right) +\zeta_2\right)  d_y \right)}.
\label{Eq:gamma_2_s5}
\end{align} 
Finally, by substituting~\eqref{Eq:gamma_1_s5} and~\eqref{Eq:gamma_2_s5} into~\eqref{Eq:gamma_s2} and then to~\eqref{Eq:Pr}, we obtain $P_r= \frac{P_{AP}}{L}$, where $L$ can be evaluated as in~\eqref{Eq:L_GC}. This concludes the proof.

\bibliographystyle{IEEEtran}
\bibliography{IEEEabrv,References}

\end{document}